\UseRawInputEncoding
\documentclass[final,5p,times,twocolumn]{elsarticle}

\usepackage{amssymb}
\usepackage{amsmath}
\usepackage{graphicx}
\usepackage{subfigure}

\usepackage{multirow}
\usepackage{algorithmic}
\usepackage{algorithm}
\usepackage{makecell}
\usepackage{xurl}

\journal{Aerospace Science and Technology}

\begin{document}

\begin{frontmatter}



\title{Non-Propulsive Payload Deployment for Efficient On-Orbit Servicing of Mega-Constellations}  

\author[inst1,inst2]{Li Zhengrui} 
\ead{lizhengrui@imech.ac.cn}
\author[inst1]{Feng Guanhua} 
\ead{fengguanhua@imech.ac.cn}
\author[inst1,inst2]{Wu Xiaokun} 
\ead{wuxiaokun0825@163.com}
\author[inst1]{Li Wenhao\corref{cor}}
\ead{liwenhao@imech.ac.cn}
\author[inst1]{Yue Yuxian\corref{cor}} 
\ead{sy1715318@buaa.edu.cn}

\cortext[cor]{Corresponding author}
\affiliation[inst1]{organization={Institute of Mechanics, Chinese Academy of Sciences},
            addressline={No.15 Beisihuanxi Road}, 
            city={Beijing},
            postcode={100190}, 
            country={China}}
            
\affiliation[inst2]{organization={School of Engineering, University of Chinese Academy of Sciences},
            addressline={19A Yuquan Road, Shijingshan District}, 
            city={Beijing},
            postcode={100049}, 
            country={China}}

\begin{abstract}



The prevailing assumption holds that on-orbit servicing (OOS) of mega-constellations is infeasible due to prohibitive fuel consumption incurred by multiple rendezvous maneuvers across vast and dispersed satellite populations. To address this challenge, a novel OOS architecture termed Non-Propulsive Payload Deployment (NPD) is proposed in this paper. Within this framework, a service spacecraft (SSc) ejects micro-payload spacecraft (PSc) into transfer orbits, after which the PSc autonomously rendezvous with target spacecraft (TSc). Since propulsion is required only for the minimal mass of the PSc, maneuvering fuel consumption is significantly reduced. This paper develops a phase-based approximation algorithm to resolve the scheduling problems arising from cumulative recoil-induced orbital perturbations. Numerical simulations for a constellation of over 100 satellites demonstrate that this algorithm reduces computation time by over 90\% while maintaining ejection velocity errors below 1\%. Further analysis yields an analytical formula for evaluating the deployment capability of the NPD system, providing planning estimates with less than 2\% error within low Earth orbit (LEO) regimes. Finally, a case study of the Starlink Gen2 constellation confirms that the NPD system consumes less than 1/50 of the propellant required by conventional methods, enabling efficient multi-plane servicing via $J_2$ perturbation.



\end{abstract}


\end{frontmatter}



\section{Introduction}

Effective space governance critically depends on on-orbit servicing (OOS), like refueling and repair, to extend spacecraft lifetimes and ensure orbital sustainability. While current OOS has proven viable for individual assets in geosynchronous orbit\cite{li_-orbit_2019,noauthor_northrop_2025,noauthor_chinas_2025,jones_china_2025},
most satellites today reside in low earth orbit mega-constellations. The prevailing assumption is that their vast numbers and dispersed nature make sustained servicing infeasible, favoring costly batch replacements over maintenance. For instance, while approximately 3,400 Starlink v1.0 and v1.5 satellites are currently operational, over 4,600 have been launched to compensate for failures and de-orbiting\cite{noauthor_jonathans_nodate}. The scale of future constellations demands a shift toward scalable, multi-target architectures to support next-generation orbital infrastructure.

To meet this scalability challenge, researchers have proposed several servicing paradigms, primarily categorized by the number of service spacecraft (SSc) and target spacecraft (TSc): one-to-many (O2M), many-to-many (M2M), and peer-to-peer (P2P). The O2M architecture, where a single SSc services multiple TSc, has been modeled as an integer programming problem for coplanar operations in circular constellations \cite{shen_optimal_2003,shen_optimal_2002} and as a path planning problem for nonplanar geosynchronous operations accounting for inclination changes \cite{alfriend_optimal_2006}. Subsequent research has focused on improving computational tractability through two-level optimization combining genetic algorithms with random search \cite{zhou_mission_2015} and enhancing algorithmic applicability by incorporating realistic orbital perturbations and comparing transfer schemes for debris removal \cite{baranov_optimal_2020,wei_scheduling_2022}. The more complex M2M architecture, a variant of O2M involving multiple SScs, frequently employs metaheuristic algorithms:genetic algorithms with auction-based initialization handle high-dimensional constraints \cite{xu_optimization_2023} and rapid repair planning \cite{yan_rapid_2025,wei_scheduling_2022}; ant colony optimization solves the tanker location-routing problem for GEO targets \cite{zhang_optimal_2019}; and particle swarm optimization optimizes multiple satellite missions \cite{daneshjou_mission_2017,zhou_optimal_2015}. The P2P paradigm, which imposes greater autonomy demands on each spacecraft as they both receive and provide services \cite{shen_peer--peer_2005}, has been shown to achieve better overall fuel efficiency than O2M in some scenarios through cooperative fuel sharing and asynchronous mixed strategies that optimize departure and return times \cite{tsiotras_comparison_2005,dutta_asynchronous_2006}. Some mission concepts also incorporate space stations (SS) as central hubs, with multi-objective planning frameworks balancing fuel cost and mission duration \cite{kong_multiobjective_2023} and optimization considering $J_2$ perturbation and window constraints \cite{zhang_multispacecraft_2014}. Related studies have focused on optimal SS deployment strategies, including location-routing for fuel tankers \cite{zhang_optimal_2019}, orbital facility location for high-altitude constellations using low-thrust Lyapunov control \cite{shimane_orbital_2024}, and precise orbit determination for fuel stations in Sun-synchronous orbit missions \cite{zhu_orbit_2020}.







These studies provide critical insights into multi-target mission planning, establishing mathematical foundations and optimization frameworks for various servicing architectures. Nonetheless, a key limitation persists across these paradigms: the need to move the entire mass of the SSc during orbital transfers. This results in fuel consumption that scales directly with the mass of the spacecraft and the frequency of maneuvers, following the fundamental rocket equation where the propellant required grows exponentially with the desired velocity change. For mega-constellations comprising hundreds or thousands of satellites, the cumulative $\Delta v $ requirements become prohibitive, making conventional servicing economically unviable and fundamentally limiting scalability. This inherent drawback motivates the exploration of a paradigm shift towards Non-Propulsive Payload Deployment (NPD). The core principle of NPD involves a SSc ejecting micro-payloads, designated as Payload Spacecraft (PSc), into precisely calculated transfer orbits; these PSc then autonomously execute the final proximity operations and rendezvous with their targets. Since only the minimal mass of the micro-payload requires propulsion for precise rendezvous, the total mission fuel consumption can be reduced by orders of magnitude compared to conventional full-spacecraft transfers.

A significant body of research on NPD systems, such as momentum-exchange tethers\cite{cartmell_generating_1998}, has focused on addressing the challenges of attitude dynamics and vibration control \cite{woodward_attitude_2025,lu_impact_2023,liu_robust_2023,aslanov_survey_2023,ismail_three_2016,feng_geomagnetic_2019}. However, a critical and less explored issue persists across all NPD implementations: the cumulative recoil from successive payload releases perturbs the SSc's orbit. Each ejection imparts an impulsive velocity change to the SSc according to momentum conservation, and these perturbations accumulate over multiple deployments, gradually altering the SSc's trajectory. This orbital drift intricately couples the parameters of each deployment event, as the initial conditions for subsequent transfers depend on the cumulative effect of all previous ejections, creating a complex scheduling problem that remains a central challenge for mission planning. Missel and Mortari proposed a path optimization method for active space debris removal using the Space Sweeper with Sling-Sat (4S) concept, where debris is captured and ejected to achieve momentum exchange \cite{missel_removing_2013,missel_path_2013}. Their approach employs evolutionary algorithms to optimize the sequence of debris interactions, maximizing removal efficiency while minimizing fuel consumption. However, their method faces fewer constraints since debris removal only requires ejecting objects into reentry trajectories, which bypasses the precise rendezvous demands inherent to OOS where payloads must achieve accurate orbital insertion and proximity operations with TSc.

To overcome this barrier, this paper develops a phase-based approximation strategy that exploits the near-circularity of the SSc's orbit after each ejection, yielding a fast planning algorithm with negligible loss of accuracy. Analysis based on this algorithm further reveals that the maximum required ejection velocity is governed by the cumulative recoil momentum, leading to a closed-form expression that eliminates the need for iteration. Finally, a comparative fuel-consumption analysis quantifies the propellant savings enabled by the NPD concept relative to traditional sequential servicing O2M.

The remainder of this paper is organized as follows. Section 2 establishes a coupled dynamics model capturing orbital perturbations from successive ejections. Section 3 presents the proposed scheduling method designed. Section 4 evaluates the accuracy and computational efficiency of this algorithm. Section 5  derives analytical expressions defining SSc deployment capability under maximum velocity constraints and analyzes the fuel efficiency of the NPD system. Finally, Section 6 provides the concluding remarks from this study.
\section{Dynamic Modeling of NPD}

\begin{figure*}[t]
\centering
\includegraphics[width=0.7\textwidth]{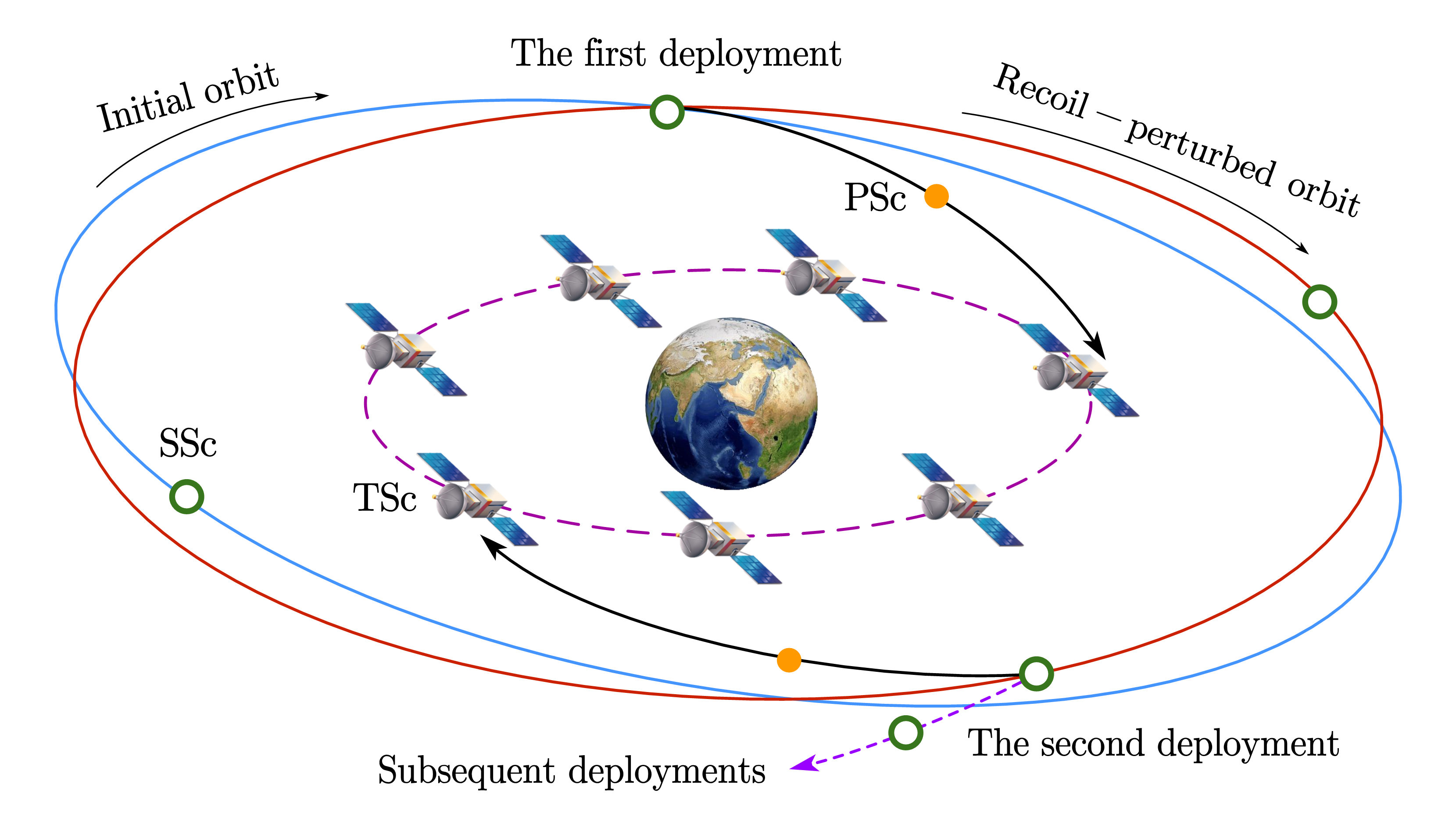}
\caption{Mission scenario}
\label{scenario}
\end{figure*}


\subsection{Mission Scenario and Assumptions}
Based on the current architecture of mega-constellations, the study examines a Walker constellation configuration where TSc reside in a defined orbital plane, The mission scenario is illustrated in Fig.\ref{scenario}.

Within this orbital plane, n TScs are uniformly distributed in phase along a circular orbit with radius $r_{t0}$ . The SSc operates in the same orbital plane, carrying n PScs. The SSc initially follows a circular orbit with radius $r_{s0}$. Through a series of successive non-propulsive ejections, each PSc is deployed to autonomously rendezvous with its designated target.

The dynamical environment assumes Earth's central gravitational field governs orbital motion, with instantaneous momentum exchange during ejections.

\subsection{Orbital Dynamics with Recoil Perturbations}

According to Keplerian mechanics, under the influence of a central gravitational field:
\begin{equation}
    \frac{\mathrm{d}^{2} \boldsymbol{r}}{\mathrm{~d} t^{2}}=-\frac{\mu}{r^{3}} \boldsymbol{r}
\end{equation}
the state transition equation for the position and velocity vectors, $\boldsymbol{r}$ and $\boldsymbol{v}$, of a spacecraft over any time interval from $t_1$ to $t_2$ is:
\begin{equation}
    \begin{bmatrix}%
     \boldsymbol{r}(t_{2})\\
    \boldsymbol{v}(t_{2})
    \end{bmatrix}=
    \begin{bmatrix}
    f(t_{2}-t_{1})& g(t_{2}-t_{1})\\
    \dot{f}(t_{2}-t_{1})  & \dot{g}(t_{2}-t_{1})
    \end{bmatrix}\begin{bmatrix}
     \boldsymbol{r}(t_{1})\\
    \boldsymbol{v}(t_{1})
    \end{bmatrix}
    \label{orb1}
\end{equation}

In Eq.(\ref{orb1}): 

\begin{equation}
\begin{array}{l}
     f(t_{2}-t_{1}) = \displaystyle{1-\frac{a}{r_1}(1-\cos\Delta E) }\\
    g(t_{2}-t_{1}) = (t_2-t_1)-\displaystyle{\sqrt{\frac{a^3}{\mu}}(\Delta E-\sin {\Delta E})}\\
    \dot{f}(t_2-t_1)  = \displaystyle{\frac{\sqrt{\mu a}\sin\Delta E}{r_1r_2}} \\
    \dot{g}(t_2-t_1) = 1-\displaystyle{\frac{a}{r_2}(1-\cos\Delta E)}
\end{array}
\end{equation}

where $r_1 = \Vert \boldsymbol{r}(t_1)\Vert$, $r_2 = \Vert \boldsymbol{r}(t_2)\Vert$, $\Vert \cdot \Vert$ denotes the magnitude of a vector and $\Delta E$ is the difference in true anomaly of the spacecraft during the time interval $t_1\leq t \leq t_2$. $\Delta E$ satisfies Kepler's equation :
\begin{equation}
    \sqrt{\frac{\mu}{a^3}}(t_2-t_1)=\Delta E-e_x\sin{\Delta E}+e_y\sin{\Delta E}
\end{equation}
and $e_x = 1-r_1/a$, $e_y = \boldsymbol{r}(t_{1}) \cdot \boldsymbol{v}(t_1)/\sqrt{\mu a}$, and $a$ is the semi-major axis. The positions and velocities of TSc and PSc, $\boldsymbol{r}_t$ and $\boldsymbol{v}_t$, and $\boldsymbol{r}_p$ and $\boldsymbol{v}_p$, respectively, all satisfy the above equations.

Since each payload deployments causes orbital change of SSc, we describe its orbital parameter as time-segmented piecewise functions. For the $i$-th TSc, the SSc ejects PSc into  transfer orbit at epoch $t_{0i}$, and PSc achieves rendezvous with its target at $t_{fi}$. Let $\boldsymbol{r}_s$ and $\boldsymbol{v}_s$ be the position and velocity vectors of SSc. Therefore,  $\boldsymbol{r}_s(t)$ and $\boldsymbol{v}_s(t)$ are difined as: 
\begin{itemize}
\item For $t$ in interval $t_{0i}<t<t_{0(i+1)}$, $\boldsymbol{r}_s(t)$ and $\boldsymbol{v}_s(t)$ are derived from Eq.(\ref{orb1})
\item Before($t_{0i}^{-}$) and after deployment($t_{0i}^{+}$), its velocity undergoes an instantaneous change during payload deployment, as described by Eq. (\ref{dv_change}):
\begin{equation}
    \begin{array}{l} 
    \boldsymbol{r}_{s}(t_{0i}^{+}) = \boldsymbol{r}_{s}(t_{0i}^{-})\\
    \boldsymbol{v}_{s}(t_{0i}^{+}) = \boldsymbol{v}_{s}(t_{0i}^{-})+\ \displaystyle{\frac{m_p}{m_s+(n-i)m_p}}(\boldsymbol{v}_{s}(t_{0i}^{-})-\boldsymbol{v}_{pi}(t_{0i}))
\end{array}
\label{dv_change}
\end{equation}
where $m_s$ and $m_p$  is the mass of SSc and PSc.
\end{itemize}

The transfer orbit of PSc is defined by its initial state vectors, $\boldsymbol{r}_{pi}(t_{0i})$ and $\boldsymbol{v}_{pi}(t_{0i})$, subject to the boundary conditions $\boldsymbol{r}_{pi}(t_{0i}) = \boldsymbol{r}_{s}(t_{0i})$ upon separation from the SSc and $\boldsymbol{r}_{pi}(t_{fi}) = \boldsymbol{r}_{ti}(t_{fi})$ upon rendezvous with the TSc. According to the solution of the Lambert problem, the positional boundary conditions and the transfer duration $t_{fi}-t_{0i}$ uniquely determine the orbital parameters of the PSc's free flight. Consequently, $\boldsymbol{v}_{pi}(t_{0i})$ is a function solely of $t_{0i}$ and $t_{fi}$, denoted as $\boldsymbol{L_{vi}}(t_{0i},t_{fi})$.

\begin{equation}
    \begin{array}{l} 
    \boldsymbol{r}_{pi}(t_{0i}) = \boldsymbol{r}_{s}(t_{0i})\\
    \boldsymbol{v}_{pi}(t_{0i}) = \boldsymbol{L_{vi}}(t_{0i},t_{fi})
\end{array}
\label{lambert}
\end{equation}

Then the magnitude of the $i$-th ejection velocity of SSc is determined by $t_{0i}$ and $t_{fi}$: 
\begin{equation}
    \Delta v_{i}=\Vert \boldsymbol{v}_{s}(t_{0i}^{-})-\boldsymbol{v}_{pi}(t_{0i})\Vert = \Vert \boldsymbol{v}_{s}(t_{0i}^{-})-\boldsymbol{L_{vi}}(t_{0i},t_{fi})\Vert = L_{i}(t_{0i},t_{fi})
    \label{orb2}
\end{equation}

Eqs. (\ref{orb1})-(\ref{orb2}) represent the orbital dynamics model of SSc. Consequently, the separation time vector $\boldsymbol{t_0} = [t_{01},t_{02},\cdots,t_{0n}]$ and rendezvous time vector $\boldsymbol{t_f} = [t_{f1},t_{f2},\cdots,t_{fn}]$, uniquely determine all key parameters of the entire NPD process, including ejection velocities, orbital transitions, and cumulative recoil effects.

\section{Deployment Scheduling under Recoil Constraints}
\subsection{Problem Formulation for Minimum Ejection Velocity}

During the deployment process, the maximum ejection velocity among the $n$ maneuvers is defined as $\Delta v_{\max}$, from the above, $\Delta v_{\max}$ could be derived from Eq. (\ref{orb2}):
\begin{equation}
    \begin{array}{l}
    \Delta v_{\max} = \max \left(\Delta v_{1}, \Delta v_{2}, \ldots, \Delta v_{n}\right)=\max \left(L\left(\boldsymbol{t}_{0}, \boldsymbol{t}_{f}\right)\right) \\
    =\max \left(L_{1}\left(t_{01}, t_{f 1}\right), L_{2}\left(t_{02}, t_{f 2}\right), \ldots, L_{n}\left(t_{0 n}, t_{fn}\right)\right)
    \end{array}
\end{equation}
Should the designed maximum ejection velocity of the SSc fall below this threshold, the mission will result in failure. 

Thus, the objective is to determine the separation and rendezvous time pairs for SSc to minimize the maximum ejection velocity $\Delta v_{\max}$. Each PSc must be deployed before its rendezvous with TSc. This logical sequence is enforced by imposing the constraint $t_{0i}<t_{fi}$.

\begin{equation}
    \begin{array}{l} 
  \text{minimize} \quad \max \left(\Delta v_{1}, \Delta v_{2}, \ldots, \Delta v_{n}\right)=\max \left(L\left(\boldsymbol{t}_{0}, \boldsymbol{t}_{f}\right)\right)\\ 
  \text{s.t.} \qquad \qquad t_{0i}<t_{fi},\forall i=1,2,\ldots,n\\
\end{array} 
\label{precise}
\end{equation}


\subsection{Greedy Algorithm for Sequential Deployment}

The greedy algorithm could be adopted to solve the optimization problem in Eq. (\ref{precise}). It sequentially minimizes the ejection velocity for each deployment. This directly reduces the orbital perturbation imparted to SSc, thereby lowering required $\Delta v_i$ for subsequent deliveries and curbing the accumulation of orbital deviations, which ultimately minimizes the overall $\Delta v_{\max}$.


\begin{algorithm}[!h]
    \caption{Greedy Algorithm for Sequential Deployment}
    \label{alg:greedy}
    \renewcommand{\algorithmicrequire}{\textbf{Input:}}
    \renewcommand{\algorithmicensure}{\textbf{Output:}}
    \begin{algorithmic}[1]
        \REQUIRE Initial orbital parameters of SSc, PSc, and TSc
        \ENSURE $\boldsymbol{t_0} = [t_{01},t_{02},\cdots,t_{0n}]$, $\boldsymbol{t_f} = [t_{f1},t_{f2},\cdots,t_{fn}]$
        \FOR{each $i \in [1,n]$}
            \STATE Solve the optimization problem for optimal $t_{0i}$ and $t_{fi}$:
            \begin{equation}
                \begin{array}{ll} 
                  \text{minimize} & \Delta v_i = L_i(t_{0i},t_{fi}) \\ 
                  \text{s.t.} & t_{0(i-1)} < t_{0i} < t_{fi}
                \end{array} 
                \label{alg:opt}
            \end{equation}
            \STATE Compute $\boldsymbol{r}_{s}(t_{0i})$ and $\boldsymbol{r}_{ti}(t_{fi})$ using Eq. (\ref{orb1})
            \STATE Determine $\boldsymbol{v}_{pi}(t_{0i})$ by solving Lambert's problem Eq. (\ref{lambert})
            \STATE Update SSc orbital parameters using momentum conservation Eq. (\ref{dv_change})
        \ENDFOR
        \RETURN Deployment time vectors $\boldsymbol{t_0}$, $\boldsymbol{t_f}$
    \end{algorithmic}
\end{algorithm}

The algorithm implementation is presented in Algorithm \ref{alg:greedy}. This approach requires the solution of $n$ two-dimensional optimization problems, and its search space grows exponentially with $n$, thereby rendering direct numerical methods impractical. In the case of LEO mega-constellations comprising over 100 TScs, the computational load becomes prohibitively large. Thus, developing a computationally efficient near-optimal solution is critical to alleviate the computational complexity.

\subsection{Phase-Based Approximation Algorithm}

Hohmann transfer is the optimal $\Delta v$ transfer between circular orbits. The angular difference between the departure and arrival positions $\theta_i = \langle \boldsymbol{r}_{s}(t_{0i}),\boldsymbol{r}_{pi}(t_{fi})\rangle$ , is $\pi$. However, the recoil momentum from deployment results in a near-circular orbit with small eccentricity for the SSc, thus deviating the PSc transfer from a standard Hohmann transfer.

The SSc's orbital eccentricity, $e_{si}$, and the deviation of the payload spacecraft's free flight segment swept angle from $\pi$, $\Delta \theta_{0i} =  |\langle \boldsymbol{r}_{s}(t_{0i}),\boldsymbol{r}_{pi}(t_{fi})\rangle-\pi|$, quantify this deviation. 

A example of Algorithm \ref{alg:greedy}, with the parameters and corresponding $e_{si}$ ,$\Delta \theta_{pi}$, is shown in Table \ref{param}. The results shows that $e_{si}$ and $\Delta \theta_{pi}$ remain small across different altitude differences, and the SSc's orbit remains close to circular.


\begin{table}[h]
\centering
\renewcommand{\arraystretch}{1.2}
\begin{tabular}{ccc}
\hline
  & Case 1 & Case 2 \\ \hline
\multirow{2}{*}{orbital altitude} &  SSc: 1050km & SSc: 1550km \\
                                   & TSc: 550km  & TSc: 550km  \\
$m_s:m_p$ & 100:1 & 100:1 \\
$n$ & 50 & 50 \\
\multirow{2}{*}{$e_{si}$} & average: 7.0e-4 & average: 2.8e-3 \\
                           & max: 1.7e-3   & max: 6.7e-3 \\
\multirow{2}{*}{$\Delta \theta_{pi}$} & average: 5.0e-3 & average: 2.4e-03 \\
                                       & max: 6.7e-2    & max: 1.3e-1 \\ \hline
\end{tabular}
\caption{Calculated orbital eccentricity and phase deviation}
\label{param}
\end{table}


Based on this, an approximate optimal transfer strategy is proposed. In this strategy, the SSc's near-circular orbit before each deployment is approximated as a circular orbit with the same semi-major axis $a_{si}$. The departure and arrival times, $t_{0i}$ and $t_{fi}$, for each transfer are then determined using the principle of an orbital transfer phase difference of $\pi$:
\begin{equation}
  \theta_i = \langle \boldsymbol{r}_{s}(t_{0i}),\boldsymbol{r}_{pi}(t_{fi})\rangle
\label{fast}
\end{equation}
    
Let $\theta_{0i}$ and $\omega_{0i}$ be the SSc's true anomaly and argument of perigee before $t_{0i}$, $\alpha_{t0}$ be the initial phase of the $i$-th TSc, the equivalent form of Eq. (\ref{fast}) is: 

\begin{equation}
    t_{fi}\sqrt{\frac{\mu}{r_{t0}^3}}+\alpha_{t0} - (\theta_{0i} + \omega_{0i}) = \pi
    \label{fast_begin}
\end{equation}

From the geometric relationship, eccentric anomaly of SSc $E_{0i}$ meet the following conditions:
\begin{equation}
    \begin{array}{cc}
        \tan \displaystyle{\frac{E_{0i}}{2}} = \displaystyle{\sqrt{\frac{1-e_{si}}{1+e_{si}}}} \tan \frac{\theta_{0i}}{2}  \\
        \displaystyle{\sqrt{\frac{\mu}{a_{si}^3}}}(t_{0i}-t_{di}) = E_{0i}-e_{si}\sin {E_{0i}}
    \end{array}
\end{equation}
where $t_{di}$ is the time of perigee passage.

For a Hohmann transfer: 
\begin{equation}
    t_{fi}-t_{0i} = \frac{\pi}{2}\sqrt{\displaystyle{\frac{(a_{si}+r_{fi})^3}{2\mu}}}
    \label{fast_end}
\end{equation}

\begin{algorithm}[!h]
    \caption{Phase-Based Approximation Algorithm}
    \label{alg:approximate}
    \renewcommand{\algorithmicrequire}{\textbf{Input:}}
    \renewcommand{\algorithmicensure}{\textbf{Output:}}
    \begin{algorithmic}[1]
        \REQUIRE Initial orbital parameters of SSc, PSc, and TSc
        \ENSURE $\boldsymbol{t_0} = [t_{01},t_{02},\cdots,t_{0n}]$, $\boldsymbol{t_f} = [t_{f1},t_{f2},\cdots,t_{fn}]$
        \FOR{each $i \in [1,n]$}
            \STATE Solve for $t_{0i}$ and $t_{fi}$ satisfying the phase-matching conditions:
            \begin{equation}
                \begin{array}{l} 
                      t_{fi}-t_{0i} = \displaystyle{\frac{\pi}{2}}\sqrt{\displaystyle{\frac{(a_{si}+r_{fi})^3}{2\mu}}}\\
                      \langle \boldsymbol{r}_{s}(t_{0i}),\boldsymbol{r}_{pi}(t_{fi})\rangle = \pi
                \end{array} 
            \end{equation}
            \STATE Compute $\boldsymbol{r}_{s}(t_{0i})$ and $\boldsymbol{r}_{ti}(t_{fi})$ using Eq. (\ref{orb1})
            \STATE Determine $\boldsymbol{v}_{pi}(t_{0i})$ by solving Lambert's problem Eq. (\ref{lambert})
            \STATE Update SSc orbital parameters using momentum conservation Eq. (\ref{dv_change})
        \ENDFOR
        \RETURN Deployment time vectors $\boldsymbol{t_0}$ and $\boldsymbol{t_f}$
    \end{algorithmic}
\end{algorithm}

Eq. (\ref{fast_begin})-(\ref{fast_end}) constitute a complete system of equations for $t_{0i}$ and $t_{fi}$, providing an implicit expression for the approximate solution to the optimization problem in Eq. (\ref{alg:opt}). Therefore, the procedure of  Algorithm \ref{alg:greedy} can be reformulated as Algorithm \ref{alg:approximate}. It simplifies the original $2n$-dimensional optimization search into solving $n$ independent systems of equations, significantly reducing computational complexity.

\section{Accuracy Evaluation for the Phase-Based Approximation Algorithm}

\begin{figure*}[htb]
\centering
\subfigure[$\sigma_1$ vs. $r_{s0}-r_{t0}$ at $m_s:m_p=200:1$]{
\includegraphics[width=0.83\textwidth]{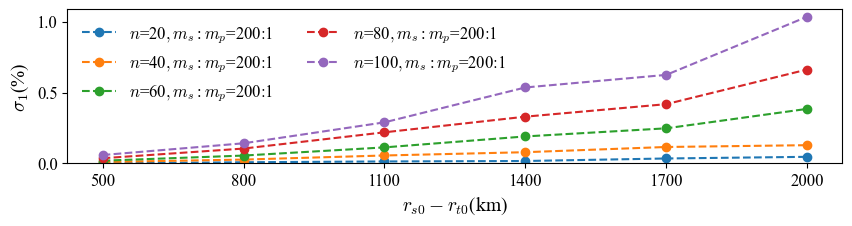}
\label{error_sum_a}
}
\subfigure[$\sigma_1$ vs. $m_s:m_p$ at $n = 50$]{
\includegraphics[width=0.83\textwidth]{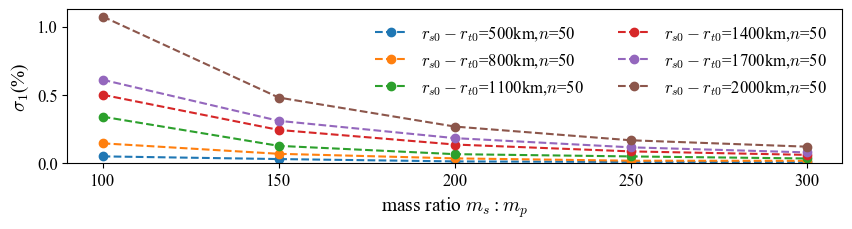}
\label{error_sum_b}
}
\subfigure[$\sigma_1$ vs. $n$ at $r_{s0}-r_{t0}=500$km]{
\includegraphics[width=0.83\textwidth]{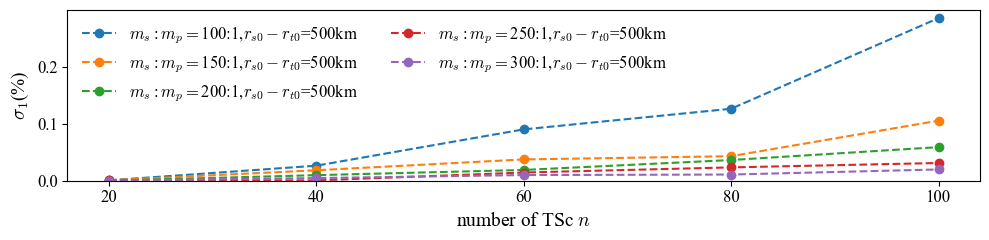}
\label{error_sum_c}
}
\caption{Approximation error ($\sigma_1$)} 
\label{error_sum}
\end{figure*}

\begin{figure*}[htbp]
\centering
\begin{minipage}[t]{0.48\textwidth}
\centering
\includegraphics[width=0.9\textwidth]{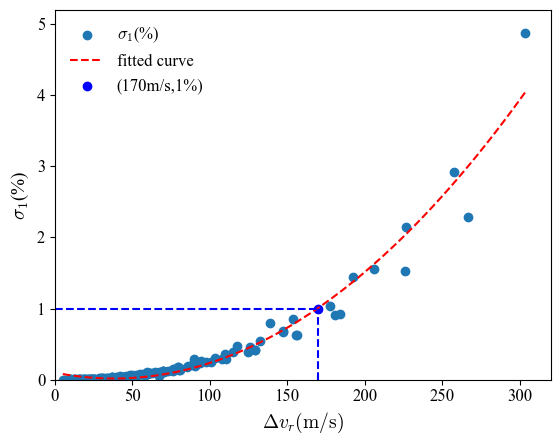}
\caption{Correlation analysis between $\sigma_1$ and $\Delta v_{r}$}
\label{error_3}
\end{minipage}
\begin{minipage}[t]{0.48\textwidth}
\centering
\includegraphics[width=1.0\textwidth]{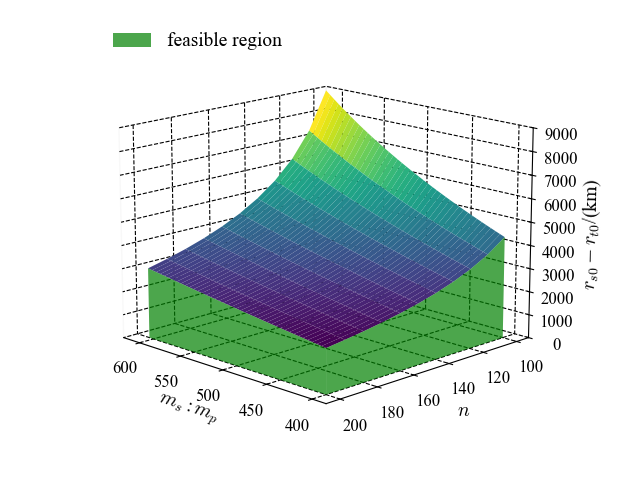}
\caption{Feasible region for approximation algorithm ($\sigma_1 < 1\%$)}
\label{sigma_1_fitted}
\end{minipage}
\end{figure*}

\subsection{Error Analysis}

In Eq.(\ref{precise}), the value of the objective function represents the maximum ejection velocity among the $n$ ejections performed by the SSc, which is denoted as $\Delta v_{\max}$. 

Accordingly, the $\Delta v_{\max}$ values obtained from Algorithm \ref{alg:greedy} and \ref{alg:approximate} are  denoted as $\Delta v_{\max1}$ and $\Delta v_{\max2}$. Simultaneously,  the relative error $\sigma_1$, between $\Delta v_{\max1}$ and $\Delta v_{\max2}$ is defined as: 

\begin{equation}
    \sigma_1 = \left|\frac{\Delta v_{\max1}-\Delta v_{\max2}}{ \Delta v_{\max1}}\right|
    \label{sigma_1}
\end{equation}

With the orbital altitude of TSc fixed at 550 km, the relative error $\sigma_1$ was calculated under the following conditions: a relative mass ratio between the SSc and TSc $m_s:m_p = 200:1$, a number of TSc $n = 50$, and a relative altitude difference between SSc and TSc $r_{s0}-r_{t0}=500$km. The corresponding results are presented in Fig. \ref{error_sum}.

A comparison of the results reveals that changes in the number of TSc, the initial altitude difference, and the mass ratio all increase the momentum change experienced by SSc, consequently leading to a higher $\sigma_1$.

By the conservation of momentum, the magnitude of the recoil velocity imparted to the SSc $\Delta v_{rj}$ upon ejection of the $j$-th PSc is:
\begin{equation}
    \Delta v_{rj} = \frac{m_p}{m_s+(n-j)m_p}\Delta v_i
\end{equation}
Consequently, the total recoil velocity change accumulated by the SSc amounts to:
\begin{equation}
    \Delta v_{r} = \displaystyle{ \sum_{j= 0}^{n} \frac{m_p}{m_s+(n-j)m_p}\Delta v_i} =\displaystyle{ \sum_{i= 0}^{n-1} \displaystyle{\frac{m_p}{m_s+im_p}}\Delta v_i}
\end{equation}
Substituting the velocity increment of Hohmann transfer yields:
\begin{equation}
    \Delta v_{r} = \left(\sqrt{\frac{\mu}{r_{t0}} \frac{2 r_{s0}}{r_{t0}+r_{s0}}}-\sqrt{\frac{\mu}{r_{t0}}}\right)\sum_{i= 0}^{n-1}\frac{m_p}{m_s+im_p}
    \label{momteum}
\end{equation}

To investigate the relationship between the cumulative recoil velocity $\Delta v_{r}$ and the relative error $\sigma_1$, we performed a parameter sweep analysis based on ranges defined in Table \ref{error_ana_table}. This table includes key parameters such as the initial orbital altitude difference $r_{s0}-r_{t0}$ the mass ratio $r_s:r_p$, and the number of TSc $n$, ensuring comprehensive and representative coverage. For each parameter combination, the corresponding $\Delta v_{r}$ and $\sigma_1$ values were computed and shown in Fig. \ref{error_3}.

\begin{table}[htbp]
\centering
\renewcommand{\arraystretch}{1.2}
\begin{tabular}{cccc}
\hline
                             & min   & max  & step  \\ \hline
 \multicolumn{4}{l}{orbital altitude of TSc: 550km} \\ 
$r_{s0}-r_{t0}$              & 600km  & 1000km      & 100km  \\
$m_s:m_p$                    & 100:1 & 500:1     & 100:1      \\
$n$                          & 20      & 100      & 20      \\ \hline
\end{tabular}
\caption{Parameter ranges for $\sigma_1$ analysis}
\label{error_ana_table}
\end{table}

A quadratic function fit to the data points yields a correlation coefficient $R^2 \geq 0.98$. This indicates that the cumulative recoil velocity, $\Delta v_{r}$ is the primary determinant of the relative error $\sigma_1$.

Assuming an acceptable $\sigma_1$ range of $0-1.0\%$, the corresponding $\Delta v_{r\max} = 170{\mathrm{m/s}}$ is obtained. Therefore, regardless of the number of TSc, as long as the cumulative recoil velocity of SSc does not exceed 170 m/s, the effective range of the Algorithm \ref{alg:approximate} can be defined by $\Delta v_{r}$ for a given mission:

\begin{equation}
    \left(\sqrt{\frac{\mu}{r_{t0}} \frac{2 r_{s0}}{r_{t0}+r_{s0}}}-\sqrt{\frac{\mu}{r_{t0}}}\right)\sum_{i= 0}^{n-1}\frac{m_p}{m_s+im_p} < \Delta v_{r\max}
    \label{simga_range}
\end{equation}

\begin{figure*}[!h]
\centering
\includegraphics[width=0.8\textwidth]{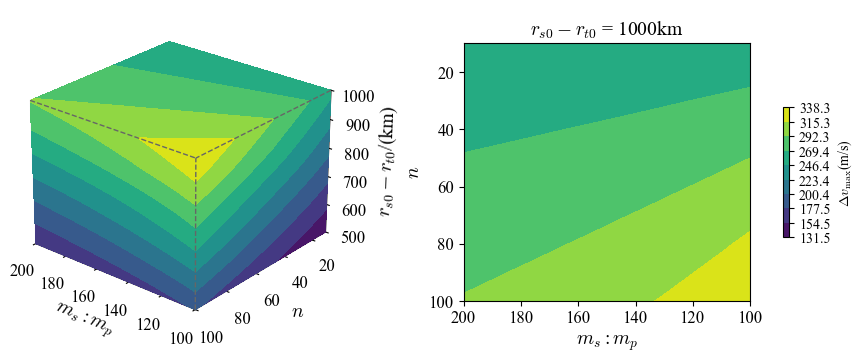}
\caption{$\Delta v_{\max}$ vs. key parameters}
\label{boundry_real}
\end{figure*}

With the TSc's orbital altitude set to $r_{t0} = 550 \mathrm{km}$, the effective ranges of $r_{s0}$, $m_s:m_p$, and $n$ derived from Eq. (\ref{simga_range}) are shown in Fig. \ref{sigma_1_fitted}.  The effective range is represented by the region below the surface, indicating that when the total mass of PSc is half the total mass of the PSc, the effective range of $r_{s0}-r_{t0}$ fully encompasses near-Earth orbital space ($> 2000\mathrm{km}$).  This demonstrates that if $\sigma_1 < 1\%$ is required, the results from Algorithm \ref{alg:approximate} closely approximate those from Algorithm \ref{alg:greedy}.  

\subsection{Computational Efficiency}

\begin{table}[htb]
    \renewcommand{\arraystretch}{1.2}
    \centering
    \begin{tabular}{cccc}
    \hline
                       & Case 1   & Case 2   & Case 3  \\ \hline
    orbital altitude   & \multicolumn{3}{c}{TSc: 550km SSc: 1050km} \\
    $m_s:m_p$              & \multicolumn{3}{c}{200:1}  \\
    $n$                  & 50      & 100     & 150    \\
    Algorithm \ref{alg:greedy}&1.90s      & 4.32s     & 7.39s   \\
    Algorithm \ref{alg:approximate}   &0.188s     & 0.37s   &0.60s    \\
    reduced            &90.1\%    & 91.4\%   &91.9\% \\ \hline
    \end{tabular}
    \caption{Comparison of computational efficiency}
    \label{time}
\end{table}

The computation times for cases in Table \ref{param} with  both algorithms are presented in Table \ref{time}, with each case run 100 times, and the average time recorded.  The simulation was conducted using an AMD Ryzen 3.8GHz processor in a Python 3.11 environment.  The results indicate that Algorithm \ref{alg:approximate} achieves an approximate optimal deployment solution with approximately 1\% error while reducing computation time by 90\%.

\begin{figure*}[htb]
\centering
\begin{minipage}[t]{0.48\textwidth}
\centering
\includegraphics[width=\textwidth]{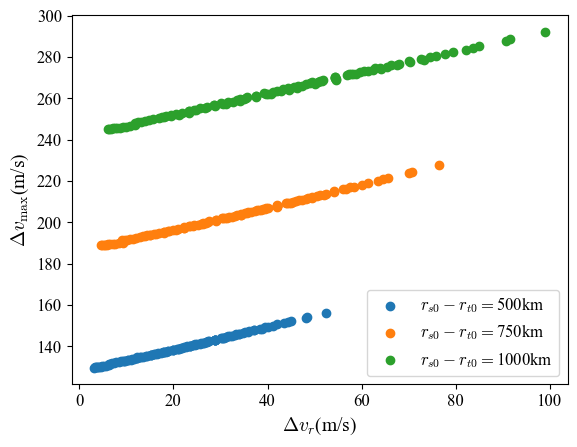}
\caption{$\Delta v_{\max}$ vs. $\Delta v_{r}$ }
\label{boundry_app1}
\end{minipage}
\begin{minipage}[t]{0.48\textwidth}
\centering
\includegraphics[width=\textwidth]{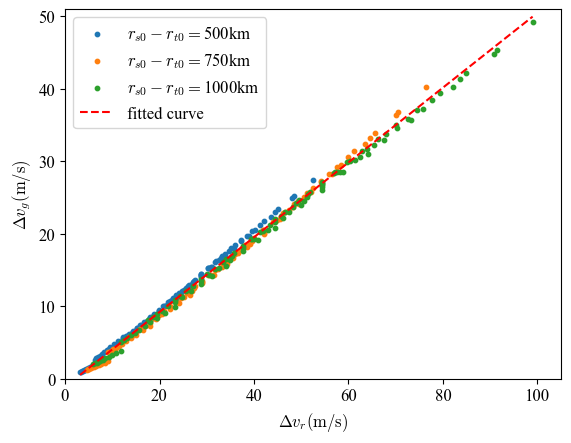}
\caption{Correlation analysis between $\Delta v_g$ and $\Delta v_{r}$}
\label{boundry_app2}
\end{minipage}
\end{figure*}

\begin{figure*}[!h]
\centering
\includegraphics[width=1.0\textwidth]{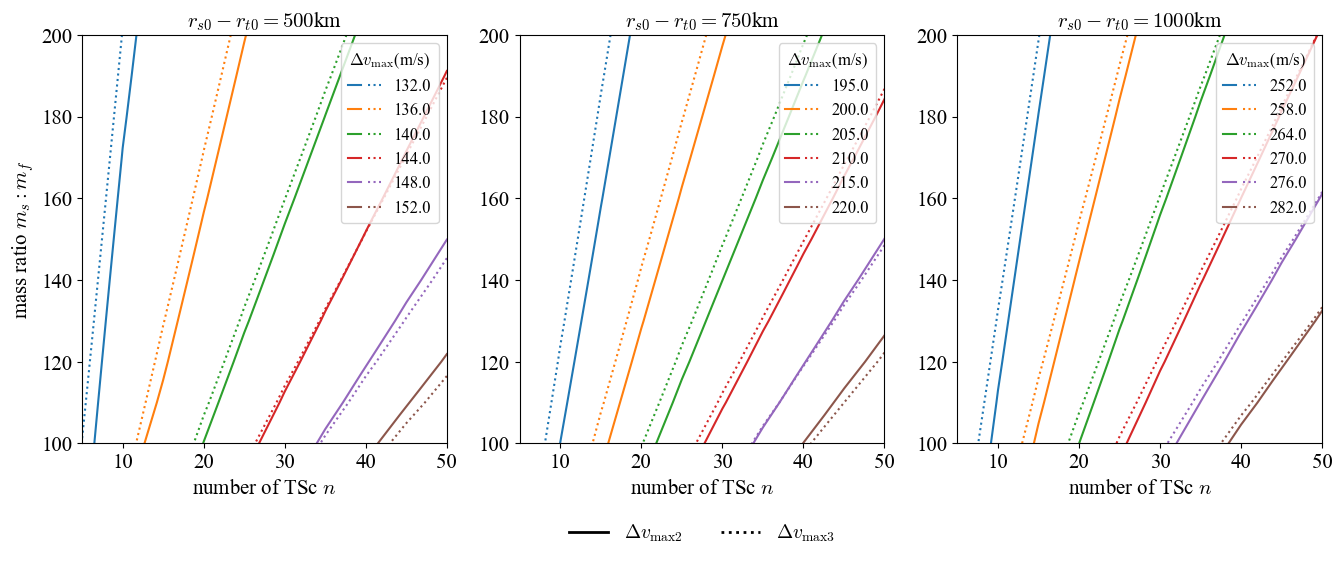}
\caption{Comparison of numerical solution and analytical formula  results}
\label{boundry_app3}
\end{figure*}

\begin{figure*}[!h]
\centering
\subfigure[$\sigma_2$ vs. $n$ at $m_s:m_p=400:1$]{
\includegraphics[width=0.46\linewidth]{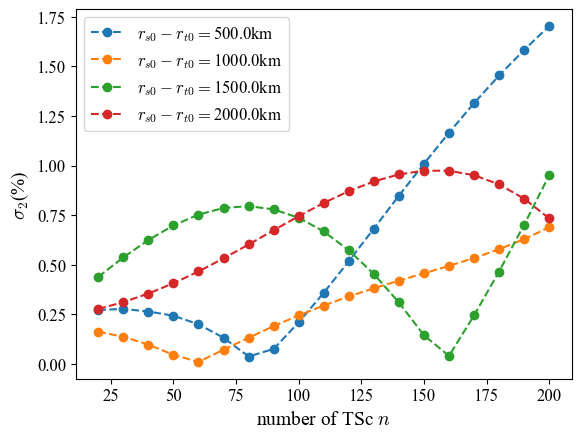}
\label{se_error1}
}
\subfigure[$\sigma_2$ vs. $r_{s0}-r_{t0}$ at $n = 50$]{
\includegraphics[width=0.45\linewidth]{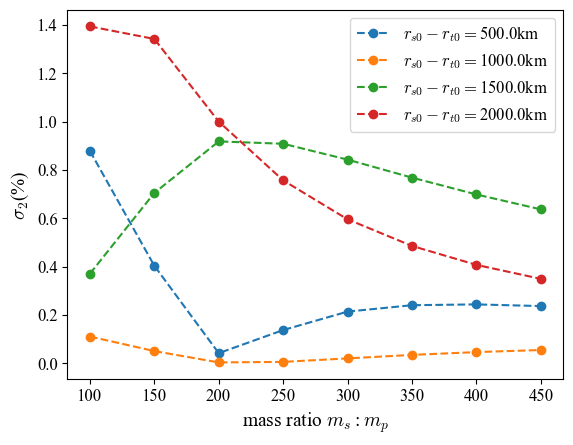}
\label{se_error2}
}
\caption{Analytical formula error ($\sigma_2$)}
\end{figure*}

\section{Deployment Capability and Fuel Efficiency Analysis}
\subsection{Deployment Capability Assessment for NPD}

$\Delta v_{\max}$ is a critical indicator of the SSc's deployment capability. Due to the orbital perturbations induced by reaction velocity, neither Eq. (\ref{alg:opt}) and Eq. (\ref{fast_begin})-(\ref{fast_end}) yields a direct analytical solution. Therefore, obtaining $\Delta v_{\max}$ for varying parameter sets necessitates numerical simulation. 

As described earlier, the three key parameters influencing reaction velocity are: the mass ratio $m_s:m_p$; the initial orbital altitude difference  $r_{s0}-r_{t0}$; and the number of TSc $n$.  Numerical simulations were performed, and the resulting $\Delta v_{\max}$ values are presented in Fig. \ref{boundry_real}.

It illustrates a monotonic relationship between $\Delta v_{\max}$ and each of the parameters: $m_s:m_p$, $r_{s0}-r_{t0}$, and $n$.  A change in any of these parameters, leading to a greater reaction velocity on SSc, results in a corresponding increase in $\Delta v_{\max}$.  Furthermore, the isolines (lines of equal value) of $\Delta v_{\max}$  are approximately linear under the condition of $r_{s0}-r_{t0} = 1000\mathrm{km}$. This indicates that for a given altitude difference, the total mass of PSc ($n\cdot m_p/m_s$) is the dominant factor determining $\Delta v_{\max}$.  Consequently, $\Delta v_{\max}$ can be considered an implicit function of $n\cdot m_p/m_s$.


Note that when $m_s >> m_p$:
\begin{equation}
    \sum_{i= 0}^{n-1}\frac{m_p}{m_s+im_p} = \sum_{i= 0}^{n-1}\frac{m_p/m_s}{1+im_p/m_s} \approx \sum_{i= 0}^{n-1}\frac{m_p}{m_s} = n\frac{m_p}{m_s}
\end{equation}
From Eq. (\ref{momteum}):
\begin{equation}
    \begin{array}{l}
        \Delta v_{r} = \displaystyle{\sum_{i= 0}^{n-1}\frac{m_p}{m_s+im_p}\left(\sqrt{\frac{\mu}{r_{t0}} \frac{2 r_{s0}}{r_{t0}+r_{s0}}}-\sqrt{\frac{\mu}{r_{t0}}}\right)}\\
        \displaystyle{\approx n\frac{m_p}{m_s} \left(\sqrt{\frac{\mu}{r_{t0}} \frac{2 r_{s0}}{r_{t0}+r_{s0}}}-\sqrt{\frac{\mu}{r_{t0}}}\right)}
    \end{array}
      \label{geq}
\end{equation}

\begin{table}[htbp]
\centering
\renewcommand{\arraystretch}{1.2}
\begin{tabular}{cccc}
\hline
                             & min   & max  & step  \\ \hline
 \multicolumn{4}{l}{orbital altitude of TSc: 550km} \\ 
$r_{s0}-r_{t0}$              & 500km  & 1000km      & 250km  \\
$m_s:m_p$                    & 100:1 & 200:1     & 10:1      \\
$n$                          & 5      & 50      & 5      \\ \hline
\end{tabular}
\caption{Parameter ranges for $\Delta v_{\max}$ analysis}
\label{dv_app_table}
\end{table}

In this case, $\Delta v_{r}$ becomes an explicit function of $n\cdot m_p/m_s$. To derive an approximate analytical formula for $\Delta v_{\max}$, its relationship with $\Delta v_{r}$ was investigated. Specifically, using the data from Table \ref{dv_app_table},  the cumulative recoil velocity  $\Delta v_{r}$ was computed for each parameter configuration based on Eq.(\ref{momteum}).  


These results are shown in Fig. \ref{boundry_app1}. It demonstrates a positive proportionality between $\Delta v_{\max}$ and $\Delta v_{r}$, for a given altitude difference.  To further isolate the effect of altitude difference, the excess velocity, $\Delta v_g$, is defined as the difference between the maximum ejection velocity and the Hohmann transfer velocity from the initial orbit:
\begin{equation}
\begin{array}{l} 
    \Delta v_g = \Delta v_{\max} - \displaystyle{\left(\sqrt{\frac{\mu}{r_{t0}} \frac{2 r_{s0}}{r_{t0}+r_{s0}}}-\sqrt{\frac{\mu}{r_{t0}}}\right)}
    \label{dvg}
\end{array}
\end{equation}

Fig. \ref{boundry_app2}  shows the relationship between $\Delta v_g$ and $\Delta v_{r}$, indicating that $\Delta v_{r}$ is the primary determinant of $\Delta v_g$ across varying altitude differences.  A fitting analysis yields the relationship: 
\begin{equation}
    \Delta v_g = 0.51\Delta v_{r}-1.08(\mathrm{m/s})
\end{equation}
It is worth noting that the linear term's coefficient is approximately 0.5.

Therefore, the velocity change resulting from orbital deviation can be approximated as $\Delta v_g = \Delta v_{r}/2$.  Substituting this into Eqs. (\ref{momteum}) and (\ref{dvg}) yields an approximate analytical expression for $\Delta v_{\max}$, denoted as \(\Delta v_{\max3}\):

\begin{equation}
\begin{array}{l} 
 \Delta v_{\max 3} = \displaystyle{\left(\sqrt{\frac{\mu}{r_{t0}} \frac{2 r_{s0}}{r_{t0}+r_{s0}}}-\sqrt{\frac{\mu}{r_{t0}}}\right) + \frac{\Delta v_{r}}{2} }\\ 
      \displaystyle{= \left(1+\frac{1}{2}\sum_{k= 0}^{n-1}\frac{m_p}{m_s+km_p}\right)\left(\sqrt{\frac{\mu}{r_{t0}} \frac{2 r_{s0}}{r_{t0}+r_{s0}}}-\sqrt{\frac{\mu}{r_{t0}}}\right)}
      \end{array}
      \label{dv_formula}
\end{equation}


Eq.(\ref{dv_formula}) implies that the SSc's deployment capability $\Delta v_{\max}$ must exceed the velocity increment required for the Hohmann transfer by at least half of the total reaction momentum change.

Similar to Fig. \ref{boundry_real}, the approximate analytical formula Eq.(\ref{dv_formula}) and Algorithm \ref{alg:approximate} were used to calculate the maximum deployment velocity corresponding to $m_s:m_p$ and $r_{s0}-r_{t0}$ for different altitude differences, as shown in Fig. \ref{boundry_app3}.

The results indicate that, under the same maximum ejection velocity constraint, the key parameter boundaries given by the analytical formula and numerical iteration are similar, and their changing trends are consistent. The error $\sigma_2$, between this analytical formula and Algorithm \ref{alg:approximate} is defined as:
\begin{equation}
    \sigma_2 = \left| \frac{\Delta v_{\max 3}-\Delta v_{\max 2}}{ \Delta v_{\max 2}}\right|
\end{equation}

With the mass ratio set to $m_s:m_p = 400:1$, $\sigma_2$ was calculated for various altitude differences, as shown in Fig. \ref{se_error1}.

With the number of TSc set to $n = 50$, $\sigma_2$ was calculated for various altitude differences, as shown in Fig. \ref{se_error2}.


Fig. \ref{se_error1} and \ref{se_error2} reveal that for deployment missions within near-Earth space (altitude differences less than 2000 km), when the total mass of PSc is less than half of the total mass of SSc, the analytical formula yields a maximum relative error of no more than 2\%. Consequently, this formula provides a reliable estimate of the maximum ejection velocity in near-Earth orbit, circumventing the need for complex iterative calculations.

\subsection{Propellant Mass Comparison of NPD and O2M}

To quantitatively assess the operational efficiency of the NPD paradigm  for on-orbit refueling, this section presents a comparative analysis between conventional servicing methods and the proposed NPD approach. Two key performance indicators are examined: the total mission completion time and the required propellant mass for maneuvers.

Referencing the orbital configuration of Starlink Gen2, the comparative study focuses on a coplanar constellation with uniformly distributed phases. Among the 120 satellites in the orbital plane, between one-fifth and all of them may require refueling; accordingly, the number of target satellites is set to range from 24 to 120. A refueling requirement of $m_{fuel}=50 \mathrm{kg}$ per target\cite{noauthor_jonathans_nodate} is assumed , with detailed simulation configurations provided in Table \ref{npd_comp_table_para}.

\begin{figure}[htb]
\centering

\includegraphics[width=0.3\textwidth]{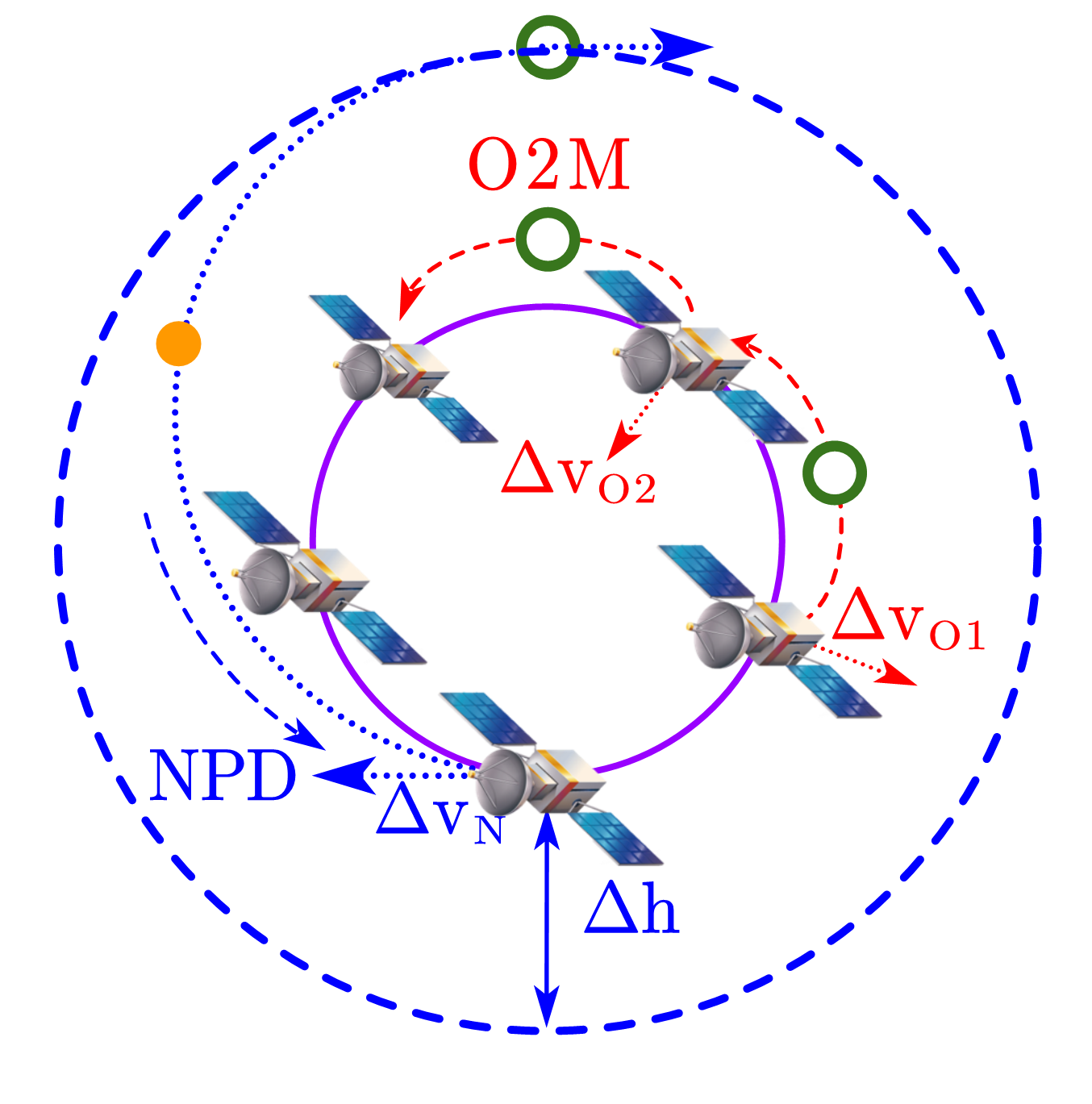}
\caption{Comparison of NPD and O2M architectures}
\label{NPD_O2M_frame}
\end{figure}

The conventional servicing paradigm assumes a sequential O2M architecture, as shown in Fig. \ref{NPD_O2M_frame},  where the SSc occupies an initial circular orbit at the same altitude as the TScs. The SSc performs successive rendezvous maneuvers to refuel TSc following the targets' orbital phase order, with each transfer duration limited by the overall schedule.  Therefore, the total velocity increment required to service a single TSc, denoted $\Delta v_O$, should be defined as the sum of the maneuver velocity $\Delta v_{O1}$ to depart from the previous TSc and the rendezvous velocity $\Delta v_{O2}$ to meet the next TSc: $\Delta v_O = \Delta v_{O1} + \Delta v_{O2}$.

\begin{figure}[htb]
\centering
\subfigure[]{
\includegraphics[width=0.2\textwidth]{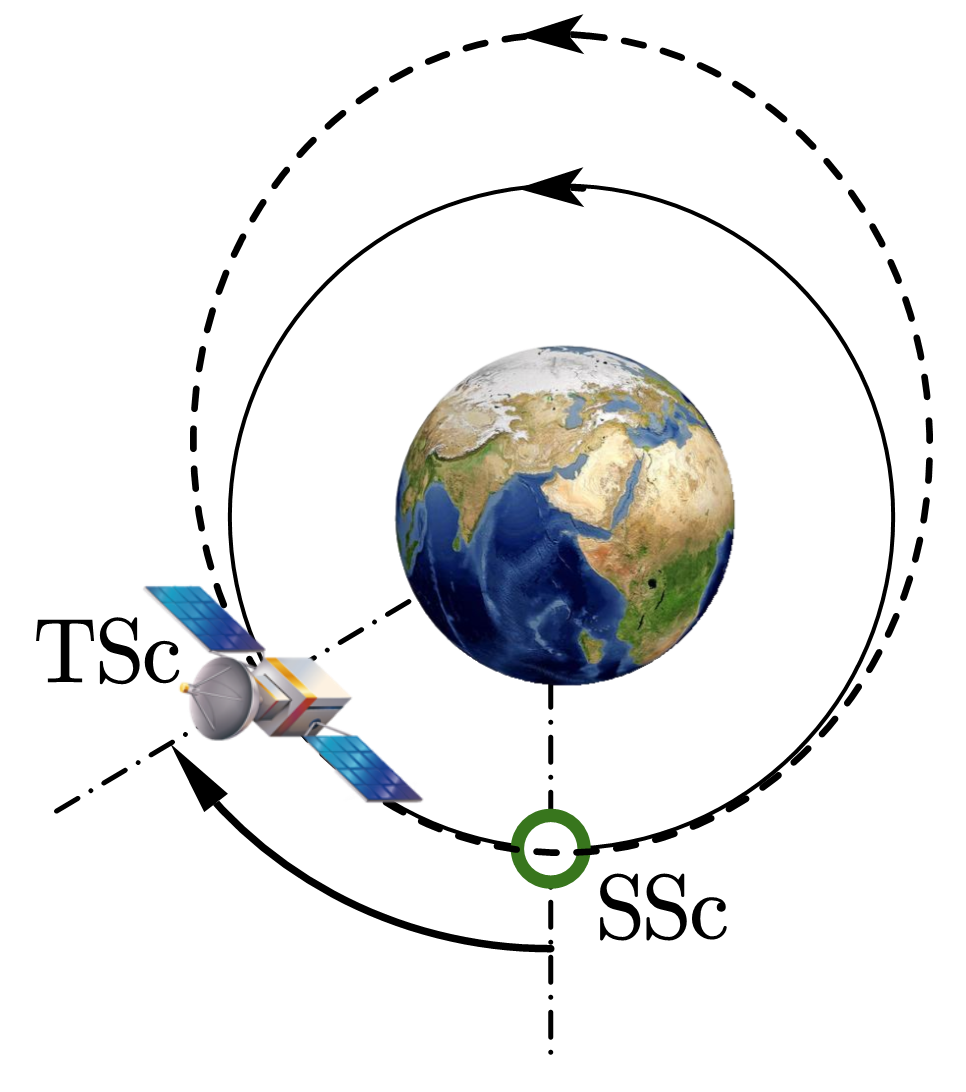}
\label{tangent_fc}
}
\subfigure[]{
\includegraphics[width=0.2\textwidth]{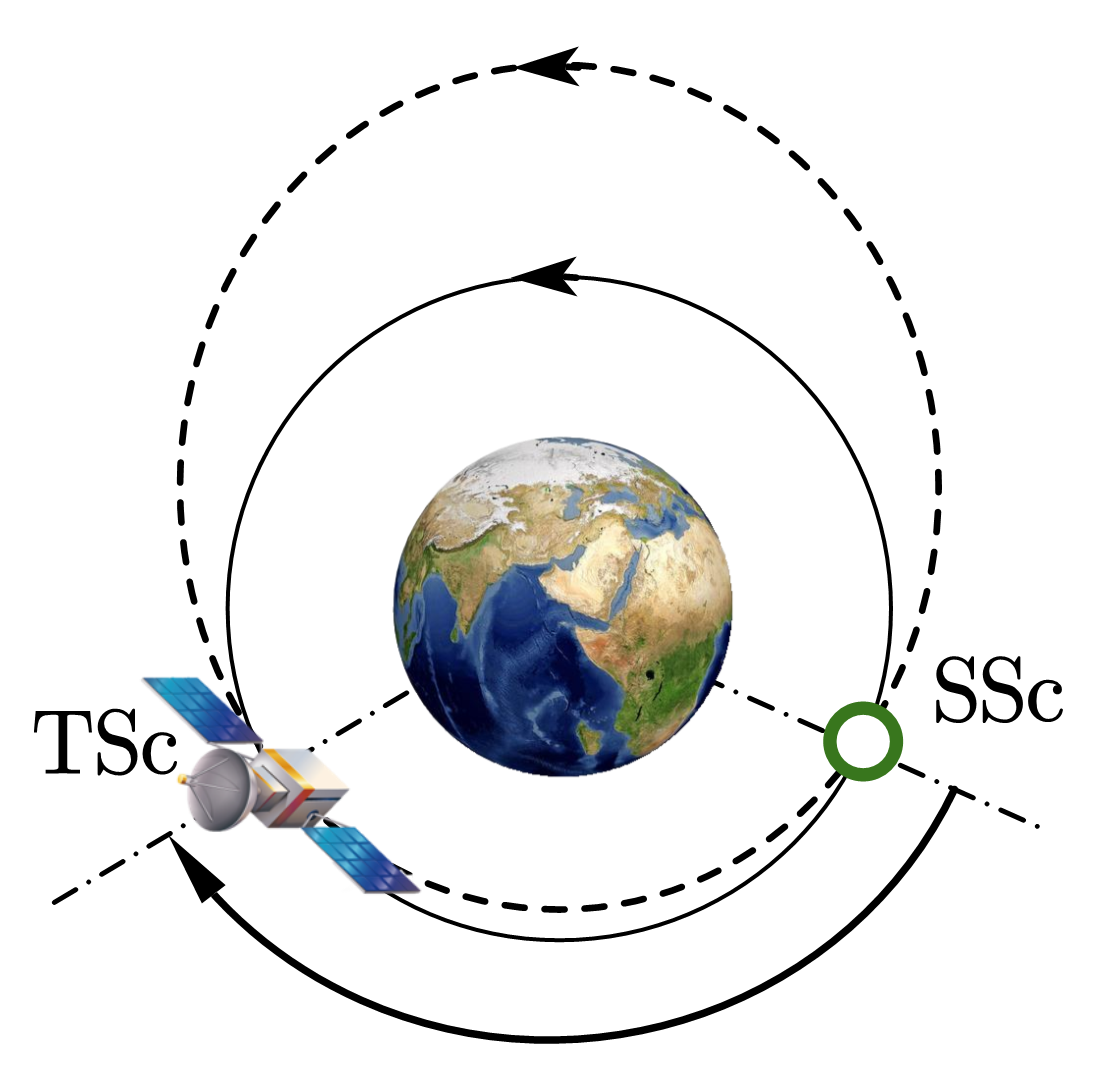}
\label{intersection_fc}
}
\caption{Phase adjustment}
\end{figure}

Given the symmetric distribution of TScs within the plane, each servicing leg requires the same $\Delta v_O$ and the same maneuver time $\Delta t_O$ under the given time window. These maneuvers are modeled as two-impulse phase adjustment transfers, manifesting either as classical tangential phasing orbits, revolving multiple times before returning to the reference altitude(Fig. \ref{tangent_fc}), or as non-tangential elliptical transfers intersecting the target orbit at two nodes(Fig. \ref{intersection_fc}). 

\begin{figure}[htb]
\centering

\includegraphics[width=0.35\textwidth]{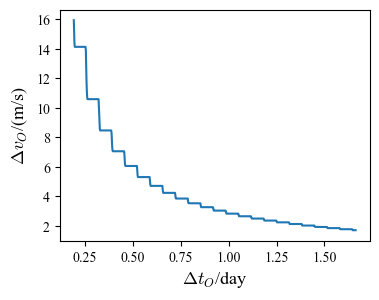}
\caption{$\Delta v_O$ for a $3^{\circ}$  phasing maneuver vs. allotted transfer time $\Delta t_O$}
\label{dv_fc}
\end{figure}

According to Shen et al.\cite{shen_optimal_2002}, employing the final coasting strategy, $\Delta v_O$ required for such phasing maneuvers exhibits a characteristic discrete step-down profile relative to $\Delta t_O$, as shown in Fig.\ref{dv_fc}.  In this figure, the $\Delta v_O$ curves correspond to a $3^{\circ}$ phasing maneuver, which represents the angular spacing between adjacent satellites in the coplanar constellation of 120 TScs.


For the NPD system, each PSc carries propellant and autonomously docks with its target TSc to transfer fuel,  propellant is consumed exclusively during the terminal rendezvous and docking between each PSc and its target TSc, shown in Fig.\ref{NPD_O2M_frame}. the time to complete the mission $t_{fn}$ is predominantly determined by the initial altitude difference between the SSc and the TSc. Simulations were performed across various altitude differentials, adopting the maximum rendezvous velocity increment $\Delta v_N$ as the reference metric to evaluate the propellant mass requirements for the PScs. 

\begin{table}[]
\begin{tabular}{cccc}
\hline
\multirow{2}{*}{}    & \multirow{2}{*}{\begin{tabular}[c]{@{}c@{}}TSc\\ parameters\end{tabular}}                                    & \multirow{2}{*}{$I_{sp}$} & \multirow{2}{*}{\begin{tabular}[c]{@{}c@{}}Mass\\      distribution\end{tabular}}                 \\
                     &                                                                                                             &                        &                                                                                                   \\ \hline
\multirow{2}{*}{O2M} & \multirow{5}{*}{\begin{tabular}[c]{@{}c@{}}altitude: 525km\\      $m_{fuel}$: 50kg\\      $n$: 24-120\end{tabular}} & \multirow{5}{*}{300s}  & \multirow{2}{*}{\begin{tabular}[c]{@{}c@{}}propellant mass: \\      1/2 of SSc \end{tabular}} \\
                     &                                                                                                             &                        &                                                                                                   \\ \cline{1-1} \cline{4-4} 
\multirow{3}{*}{NPD} &                                                                                                             &                        & propellant mass:                                                                                 \\
                     &                                                                                                             &                        & 1/4 of SSc+TScs                                                                              \\
                     &                                                                                                             &                        & $m_s:m_p$=120:1                                                                                   \\ \hline
\end{tabular}
\caption{Simulation parameters of the SSc}
\label{npd_comp_table_para}
\end{table}

After obtaining $\Delta v_O$ and $\Delta v_N$, the propellant mass under the two methods can be determined based on their respective mass distribution parameters in Table \ref{npd_comp_table_para}. For the O2M architecture, the SSc is assumed to allocate half of its total mass to propellant (including both the fuel transferred to TSc and that consumed for rendezvous maneuvers). In contrast, for the NPD architecture, to reserve more structural mass for the deployment mechanism, the SSc is designed to have a mass equal to the total mass of the TSc it carries, while each PSc allocates only half of its mass to propellant. Consequently, for the combined system of the SSc and its payloads, only one quarter of the total mass is dedicated to propellant. 

\begin{figure}[htb]
\centering

\includegraphics[width=0.5\textwidth]{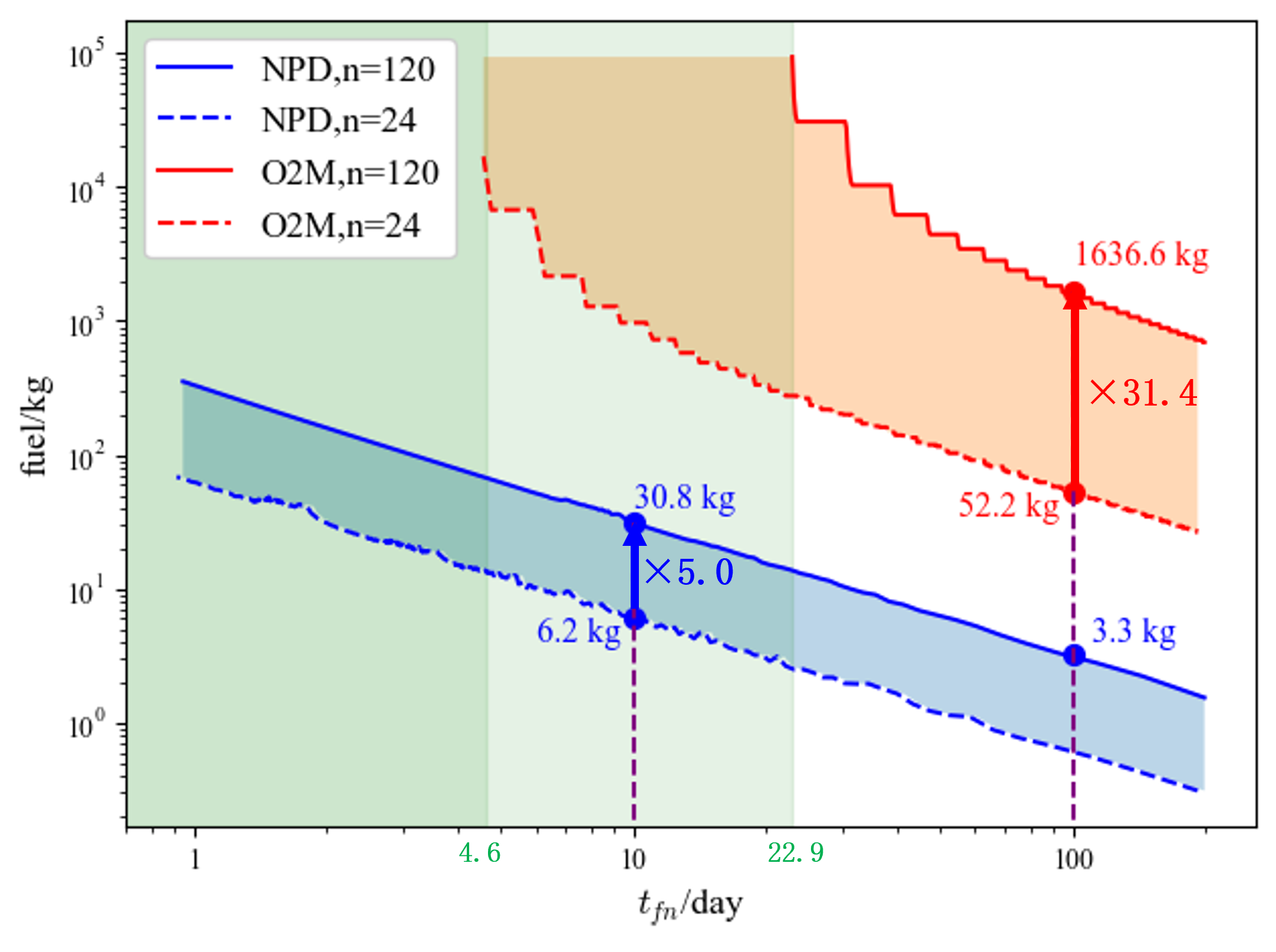}
\caption{Propellant mass comparison for NPD and O2M}
\label{dv_fc_compare}
\end{figure}

Fig. \ref{dv_fc_compare} shows the comparison of propellant mass consumed for maneuvering between the two methods. For a constellation of 120 target satellites, the O2M approach becomes physically impossible when the transfer duration is shorter than 22.9 days, and the same applies to smaller fleets: servicing 24 satellites becomes infeasible within 4.6 days. By contrast, NPD remains viable under these aggressive timelines and consistently consumes dramatically less propellant. For a 10-day mission with 120 targets, NPD expends merely 30.8 kg of maneuvering propellant-less than 1/50 of the 1636.6 kg required by O2M for a 100-day mission, while completing the task in 1/10 of the time. Moreover, NPD's propellant consumption is remarkably insensitive to the number of targets; scaling from 24 to 120 TSc increases the NPD propellant mass by only a factor of 5, whereas the O2M requirement escalates by more than 30-fold. These results demonstrate that NPD enables rapid-response servicing far beyond the capability of conventional O2M methods.

\begin{figure}[htb]
\centering

\includegraphics[width=0.4\textwidth]{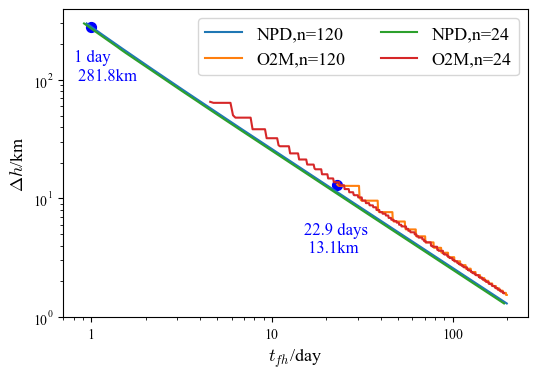}
\caption{$\Delta h$ vs. $t_{fh}$ for NPD and O2M}
\label{dv_fc_compare_dh}
\end{figure}

\begin{figure}[htb]
\centering

\includegraphics[width=0.35\textwidth]{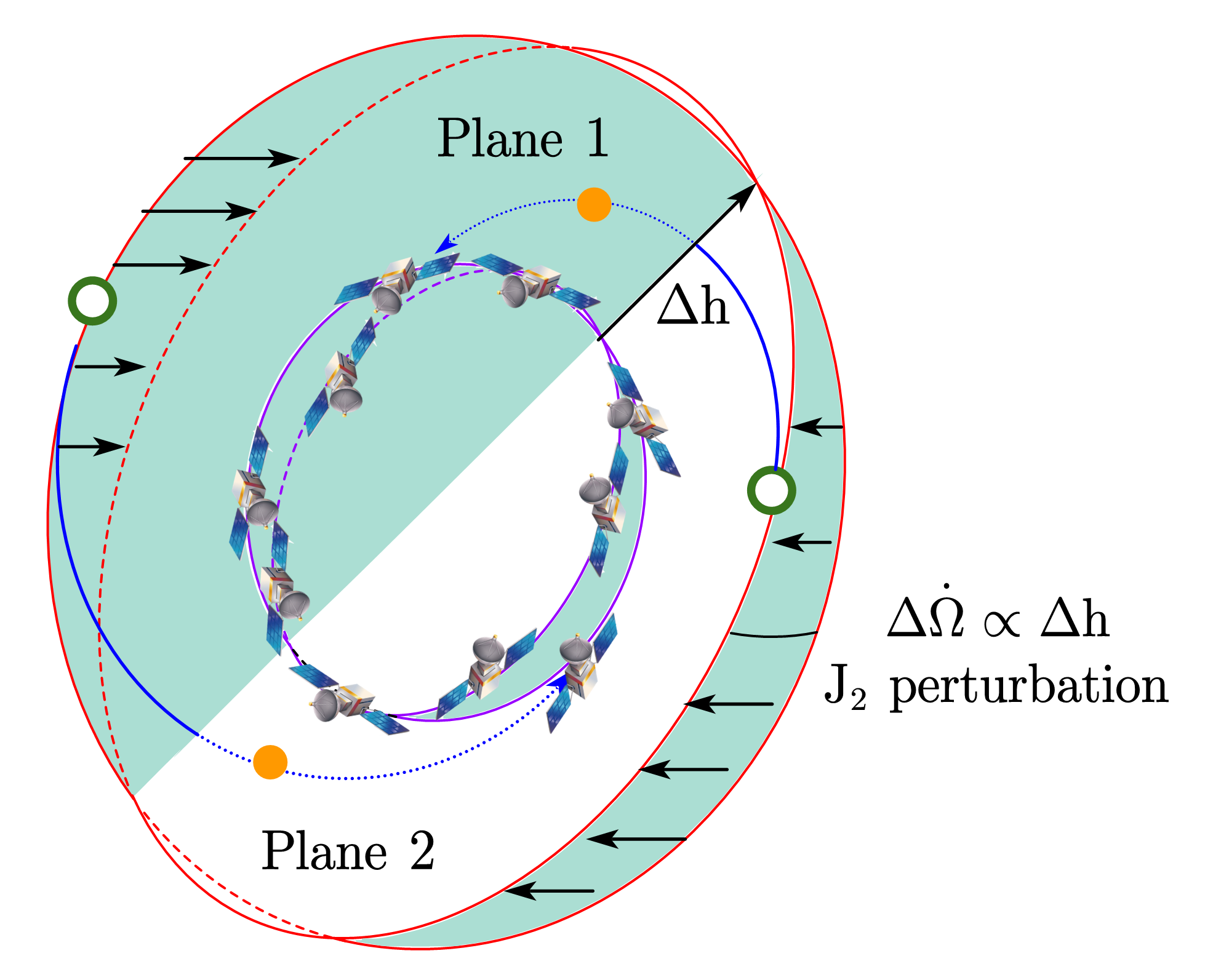}
\caption{Orbital plane drift of the SSc in the NPD system}
\label{dv_fc_compare_inter_plane}
\end{figure}

To further analyze the SSc orbits under different mission completion time constraints $t_{fn}$, we compare the orbital altitude difference $\Delta h$ for both architectures. For NPD, $\Delta h$  is defined as $r_{s0}-r_{t0}$; for O2M, $\Delta h$ refers to the difference between the semi-major axis of the SSc's transfer orbit and $r_{t0}$. The relationship between $\Delta h$ and $t_{fn}$ is presented in Fig \ref{dv_fc_compare_dh}.

Taking $n=120$ as an example, the SSc in the NPD method can achieve a delivery time as short as one day by operating in a higher orbit (with an $\Delta h$ of 281.8 km), whereas the O2M method fails to do so. The minimum achievable mission time for O2M is 22.9 days, corresponding to an altitude difference of 13.1 km. 

A larger altitude difference also leads to a faster relative nodal precession rate $\Delta \dot{\Omega}$ with respect to the TSc's orbital plane under $J_2$ perturbation, enabling the SSc to complete a deployment in a shorter time and subsequently drift to the next TSc's orbital plane for further deployments, as shown in Fig.\ref{dv_fc_compare_inter_plane}. Given the extremely small eccentricity of the SSc's orbit, $\Delta \dot{\Omega}$ can be expressed as:
\begin{equation}
    \Delta \dot{\Omega} = -1.5J_2\cos{i}\left(\sqrt{\frac{\mu}{r_{s}^3}}\left(\frac{R_E}{r_{s}}\right)^2-\sqrt{\frac{\mu}{r_{t0}^3}}\left(\frac{R_E}{r_{t0}}\right)^2\right)
\end{equation}
where $R_E$ is the Earth's radius, $J_2$ is the Earth's second-degree zonal harmonic coefficient, $i=53^{\circ}$, is the orbital inclination of the target constellation, and $r_s = r_{t0}+\Delta h$, is the SSc's orbit radius.

Assume the target constellation consists of 56 orbital planes. Then the difference in right ascension of the ascending node between adjacent planes is $\Delta \Omega = 2\pi/56$. For NPD, the SSc operating at an altitude difference of 281.8 km relative to the TScs requires only 10.8 days to transfer to an adjacent orbital plane, whereas O2M takes as many as 213.6 days. If O2M were to maneuver to a higher orbit to achieve the required nodal precession drift, it would consume significantly more propellant, further highlighting the fuel efficiency advantage of the NPD architecture.

\section{Conclusion}
This paper has proposed the NPD architecture to address the infeasibility of conventional on-orbit servicing for mega-constellations, which arises from prohibitive fuel consumption.  The core findings and contributions are summarized as follows:
\begin{enumerate}
    \item The proposed phase-based approximation algorithm significantly reduces computational complexity while maintaining high accuracy. By transforming the high-dimensional optimization problem into tractable subproblems based on Hohmann transfer principles, it enables near-real-time mission planning for large-scale constellations, saving over 90\% of computation time with less than 1\% error in ejection velocity.
    \item An analytical expression for the maximum required ejection velocity is derived, revealing that it is directly governed by the cumulative recoil velocity. Specifically, the maximum ejection velocity equals the standard Hohmann transfer velocity plus half of the total recoil velocity accumulated from all previous ejections. This formula enables rapid mission assessment without intensive simulation, and its error remains below 2\% within near-Earth orbital regimes.
    \item The NPD architecture exhibits superior fuel efficiency, with propellant consumption remaining largely insensitive to the number of TSc. Numerical simulations show that the maneuvering propellant consumed by the NPD system is less than 1/50 of that required by traditional O2M approaches. By operating at a higher orbit, the SSc reduces mission completion time, and the increased altitude difference facilitates nodal precession, enabling sequential servicing across multiple orbital planes.
\end{enumerate}


\section* {CRediT Authorship Contribution Statement}
\textbf{Li Zhengrui}: Data Curation, Formal analysis, Methodology, Writing - Original Draft. 
\textbf{Feng Guanhua}: Visualization, Writing - Review \& Editing. 
\textbf{Wu Xiaokun}: Validation, Visualization. 
\textbf{Li Wenhao}: Conceptualization, Project administration, Supervision.  
\textbf{Yue Yuxian}: Formal analysis, Writing - Review \& Editing

\section*{Declaration of Competing Interest}
The authors declare that they have no known competing financial interests or personal relationships that could have appeared to influence the work reported in this paper.

\section* {Acknowledgements}
This work is supported by the Chinese Academy of Sciences Project for Young Scientists in Basic Research (YSBR-107), Strategic Priority Research Program of Chinese Academy of Sciences(Grant No. XDA0470101, No. XDA0470101)

 \bibliographystyle{elsarticle-num} 
 \bibliography{bibitem}






\end{document}